\documentclass[conference]{IEEEtran}
\IEEEoverridecommandlockouts

\usepackage{cite}
\usepackage{amsmath,amssymb,amsfonts}
\usepackage{algorithmic}
\usepackage{graphicx}
\usepackage{textcomp}
\usepackage{xcolor}
\usepackage{booktabs}

\graphicspath{{figures/}}

\makeatletter
\def\ps@IEEEtitlepagestyle{%
  \def\@oddhead{}\def\@evenhead{}%
  \def\@oddfoot{\parbox[b]{\textwidth}{\centering\scriptsize
    \copyright~2026 IEEE. Personal use of this material is permitted. Permission from IEEE must be obtained for all other uses, in any current or future media, including reprinting/republishing this material for advertising or promotional purposes, creating new collective works, for resale or redistribution to servers or lists, or reuse of any copyrighted component of this work in other works.}}%
  \def\@evenfoot{}}
\makeatother

\begin{document}

\title{Schedule Repair for DAG Workflows under Link Disruptions}

\author{
\IEEEauthorblockN{Mohammadali Khodabandehlou\IEEEauthorrefmark{1}, Jared Coleman\IEEEauthorrefmark{2}, Bhaskar Krishnamachari\IEEEauthorrefmark{1}, Kevin Chan\IEEEauthorrefmark{3}}
\IEEEauthorblockA{\IEEEauthorrefmark{1}Dept.\ of Electrical and Computer Engineering, University of Southern California, Los Angeles, USA}
\IEEEauthorblockA{\IEEEauthorrefmark{2}Dept.\ of Computer Science, Loyola Marymount University, Los Angeles, USA}
\IEEEauthorblockA{\IEEEauthorrefmark{3}U.S. DEVCOM Army Research Laboratory, Adelphi, USA}
}

\maketitle

\begin{abstract}
Schedules for directed acyclic graph (DAG) workflows in networked IoT systems are typically computed assuming a static or generally stable network. In contested and adversarial environments, this assumption is not valid. Links degrade and fail due to mobility, interference, and jamming.
We study \emph{schedule repair}: when a link disruption invalidates part of a schedule, how much of it should be rescheduled?
We introduce a spectrum of repair policies that vary in repair scope, how much of the pending schedule each may move: wait out the disruption, reroute data around it, reschedule only the affected tasks locally, or reschedule all pending tasks globally. We evaluate each against an oracle and charge every repair a decision latency proportional to the extent to which it moves.
Across 100 workload instances spanning synthetic task graphs, RIoTBench pipelines, and WfCommons scientific workflows, each run at five communication-to-computation ratios (CCRs) and disrupted by processes with deliberately different correlation structure, we find that no single scope wins: rerouting nearly erases isolated failures that cost waiting 30\%, global repair comes within 4\% of the oracle under jamming blackouts, waiting is favored under memoryless link flapping for larger and communication-heavy workloads (the scheduling analog of route-flap damping), self-healing mobility outages reward patience over reaction, and accounting for repair latency erodes large scopes first.
We conclude that the scope of the repair should be adapted to the disruption process and the repair cost, rather than fixed by the scheduler.
\end{abstract}

\begin{IEEEkeywords}
task graph scheduling, link failure, schedule repair, rescheduling, contested environments, edge computing
\end{IEEEkeywords}

\section{Introduction}
\label{sec:introduction}

Mission-critical IoT applications such as sensor fusion pipelines, stream analytics, and distributed inference are structured as task graphs (directed acyclic graphs, DAGs) executing across heterogeneous networked compute~\cite{shukla2017riotbench, ghosh2021jupiter}. Their schedules are produced by list heuristics such as Heterogeneous Earliest Finish Time (HEFT)~\cite{topcuoglu2002heft} under the assumption that the network holds still, which an adversarial environment rarely honors.
In contested and adversarial settings, the network is what changes: links degrade and fail under mobility, interference, and deliberate jamming.
A schedule planned against a static network is then partially invalid, but it cannot simply be discarded. Tasks that started are committed, their outputs sit on particular nodes, and every transfer in flight is a sunk or lost cost. What remains is a decision about how much of the pending schedule a disruption should tear up.

This paper studies that decision as a spectrum of repair policies of increasing \emph{repair scope}, evaluated against an oracle that knows the disruption trace in advance.
Prior work examined the analogous knob in workload dynamics. As new task graphs arrive over time, the scheduler must decide how many previously placed tasks each arrival may displace, trading off makespan against fairness and overhead~\cite{khodabandehlou2025preemption}.
In this work, the workload is fixed, and the network changes.

We argue and show empirically that no single point on this spectrum is correct, because the answer depends on the structure of the disruption process and the price of repair.
We make four contributions:
\begin{itemize}
    \item We define the repair ladder with precise, policy-independent disruption semantics, including a repair cost model that charges each plan a decision latency proportional to the number of tasks it moves, while the old schedule continues executing.
    \item We build four disruption generators with deliberately different correlation structures.
    \item We implement all of it in the ncsim simulator~\cite{krishnamachari2026ncsim} with deterministic, seed-reproducible runs.
    \item Across 500 variants (100 workload instances spanning synthetic structures, RIoTBench pipelines~\cite{shukla2017riotbench}, and WfCommons scientific workflows~\cite{coleman2023wfchef}, each at five CCRs), we show that repair scope must match both the disruption structure and the communication regime.
\end{itemize}

Our findings support a ladder rather than a single winner.
Rerouting nearly erases isolated failures that would otherwise cost 30\% waiting time, and global repair comes within 4\% of the oracle under jamming blackouts.
Waiting is favored under memoryless link flapping for larger and communication-heavy workloads; self-healing mobility outages reward patience over reaction, and accounting for repair latency erodes large scopes first.

\section{Related Work}
\label{sec:related-work}

\subsection{Fault-Tolerant Workflow Scheduling}

Fault tolerance in DAG and workflow scheduling has traditionally meant redundancy. Tasks are replicated across resources or checkpointed for resubmission, so a failure costs a re-execution rather than a lost workflow~\cite{setlur2020faulttolerant}.
Mei et al.~\cite{mei2015ftdr} reschedule tasks suspended by resource failures and tolerate an arbitrary number of them.
Across this literature, the failing component is a \emph{processor} (a VM, a cluster node, a grid site) in a wired datacenter or grid, and the response is provisioned before the failure through spare capacity.
Our setting inverts both assumptions: the failing component is a link in a wireless network with no spare capacity to provision, and the response is reactive, with the schedule itself repaired after the disruption is observed.

\subsection{Reactive Rescheduling and Failure-Resilient Edge Scheduling}

A second strand asks not how to provision for failure but \emph{when to reschedule}.
Sakellariou and Zhao~\cite{sakellariou2004rescheduling} reschedule a grid workflow only at selected points where measured delay exceeds the schedule's slack, achieving most of the benefit at a fraction of the cost.
The trigger, however, is drift in cost estimates for a wired grid; the question of how the rescheduling scope should relate to the location of a discrete network event does not arise.
Recent work adapts classic list heuristics, HEFT included, to online execution under runtime uncertainty~\cite{chamorro2025online}; the trigger is again deviation from estimates rather than a discrete, located network event.
Closest to our work, Cai et al.~\cite{cai2021datr} reschedule dependent tasks when an edge server fails, showing that classical DAG scheduling breaks at the edge without rescheduling and contention awareness.
Their repair is a single fixed policy for node failures; we study link failures and make the scope of repair the object of comparison.
The same trade-off was studied under \emph{workload} dynamics: when new task graphs arrive over time, partially preemptive schedulers that replan only the most recent $K$ graphs achieve most of the makespan and utilization gains of full preemption at far lower fairness and runtime costs~\cite{khodabandehlou2025preemption, khodabandehlou2025sensys}.
This paper asks the complementary question for \emph{platform} dynamics: when the network changes while the workload remains fixed, how much of the schedule should move?

\subsection{Repair in Networks}

The vocabulary of local versus global repair originates in networking. Ad hoc On-Demand Distance Vector (AODV) routing~\cite{perkins1999aodv} repairs a broken route locally at the point of failure rather than rediscovering it end-to-end; Multiprotocol Label Switching (MPLS) fast reroute~\cite{rfc4090} pre-installs local detours because global re-optimization is too slow; and Border Gateway Protocol (BGP) route flap damping~\cite{rfc2439} suppresses reaction to links that fail and recover too quickly to be worth chasing.
Pozo et al.~\cite{pozo2021shp} lift this question from routes to schedules: after a link failure in a time-triggered network, their self-healing protocol repairs the transmission schedule online, with only small local synthesis problems, in milliseconds. In contrast, full rescheduling takes orders of magnitude longer. Their schedules, however, allocate time-division multiple access (TDMA) slots to frames; computation does not move.
We port the local-versus-global repair question to \emph{compute} schedules, where repairing means rescheduling tasks on heterogeneous nodes, with makespan rather than frame-set schedulability as the objective.

\subsection{Scheduling in Dynamic Wireless Environments}

Task offloading in mobile edge computing directly confronts link dynamics, and recent work is explicitly adversarial: Asemian et al.~\cite{asemian2025jamming} combine transmission diversity with jamming-aware scheduling to keep offloading reliable even under attack.
This line, like the larger learning-based offloading literature, optimizes placement for arriving, mostly independent tasks under channel uncertainty; there is no standing multi-task schedule to repair.
Bursty link behavior has been modeled using two-state Markov chains since Gilbert and Elliott~\cite{gilbert1960, elliott1963}, and mobility-induced link lifetimes have been analytically characterized~\cite{nayebi2012linklifetime}; we use both as trace generators with controlled statistical structure.
Decentralized dispatchers such as Jupiter execute task graphs on networked edge devices and must already cope with degraded links in practice~\cite{ghosh2021jupiter}, but do not study repair policy; simulators for networked compute have likewise treated the network as stable~\cite{krishnamachari2026ncsim}.

To our knowledge, no prior work compares repair scopes for compute task graph schedules under link-level disruptions, nor asks how the right scope depends on the statistical structure of the disruption process.

\section{System Model}
\label{sec:system-model}

\textbf{Platform.}
A network is a set of compute nodes $V$, each with speed $s(v)$, placed in the plane and connected by directed radio links with bandwidth $b(u,v)$.
A bidirectional radio link is a pair of directions, and a wireless failure takes out the pair.
Nodes execute one task at a time; concurrent transfers on a link share its bandwidth fairly.

\textbf{Workload.}
A workload is a task graph $G = (T, D)$.
Task $t \in T$ has compute cost $c(t)$ and executes on node $v$ in time
\begin{equation}
    t_{\mathrm{exec}}(t, v) = c(t) / s(v).
    \label{eq:exec}
\end{equation}
A dependency $(t, t') \in D$ carries $d(t, t')$ bytes; with $t$ placed on $u$ and $t'$ on $v$, delivery over the link (or multi-hop route) between them takes
\begin{equation}
    t_{\mathrm{xfer}}(t, t') = d(t, t') / b(u, v),
    \label{eq:xfer}
\end{equation}
and is instantaneous when $u = v$.
A scheduler assigns tasks to nodes; execution is event-driven, so start times emerge from when inputs arrive and nodes free up rather than being fixed in advance.

\textbf{Disruption semantics.}
The platform is dynamic: a \emph{disruption trace} is a time-ordered sequence of link events, \textsc{down} (the link becomes unusable), \textsc{degrade} (it stays up at a fraction of its bandwidth), and \textsc{recovery}.
Four rules fix the semantics for every policy we compare: a transfer in flight on a failed link is aborted and later retransmitted in full; a transfer with no usable route stalls until recovery or repair provides one; tasks that started are \emph{committed} and never move, so only pending tasks may be replanned; and a workload that cannot finish (e.g., a permanently severed route) is reported as a mission failure rather than folded into makespan.

\textbf{Repair is not free.}
When a policy computes a new placement at time $\tau$ moving $k$ tasks, the plan takes effect only at
\begin{equation}
    \tau' = \tau + k \, \delta,
    \label{eq:repair-cost}
\end{equation}
where $\delta$ is the per-task decision latency.
Policies may replan on any link event, including recoveries; the old schedule keeps executing during the window; whatever commits in the meantime wins; and a newer plan supersedes a pending one.
State observation and plan dissemination happen out of band, so $\delta$ can absorb control-plane latency as well as computation time.
The number of moved tasks $k$ is the plan's realized repair scope, so Eq.~\eqref{eq:repair-cost} prices scope directly: global repair moves many tasks and pays a long delay, local repair moves few, and waiting pays nothing.
The model incurs decision-latency costs but not the dissemination of new assignments over the same disrupted network, which we leave to future work on decentralized repair.

\subsection{The Repair Ladder}
\label{sec:repair-ladder}

When a disruption invalidates part of the schedule, the policies we compare differ in exactly one thing, their \emph{repair scope}: the set of pending tasks the policy may move in response.
Ordering the policies by increasing scope yields the \emph{repair ladder}:

\begin{itemize}
    \item \textbf{Wait} (L0): nothing moves. Stalled transfers retry when the link recovers.
    \item \textbf{Reroute} (L1): data detours around the failure over multi-hop routes; no task moves.
    \item \textbf{Local repair} (L2): replan only the tasks the disruption reached (those with an undelivered input whose endpoints straddle the changed link) plus their pending descendants, since moving a task moves where its output lands; every other pending task stays pinned.
    \item \textbf{Global repair} (L3): replan every pending task against the disrupted network.
    \item \textbf{Oracle}: a foresight reference, not a globally optimal scheduler; it sees the complete trace but still places with HEFT. The trace makes the network piecewise-static; we execute each phase's HEFT placement against it, add the free-repair global run, and take the best completed run.
    
\end{itemize}

Both rescheduling policies delegate placement to the same list scheduler (HEFT~\cite{topcuoglu2002heft}) with committed tasks pinned at their actual execution windows, so the policies differ only in scope, never in placement quality.
A moved task re-acquires inputs already delivered to its old node, and in-flight deliveries toward it are aborted.

\section{Disruption Models}
\label{sec:disruption-models}

\begin{figure*}[t]
    \centering
    \includegraphics[width=0.9\textwidth]{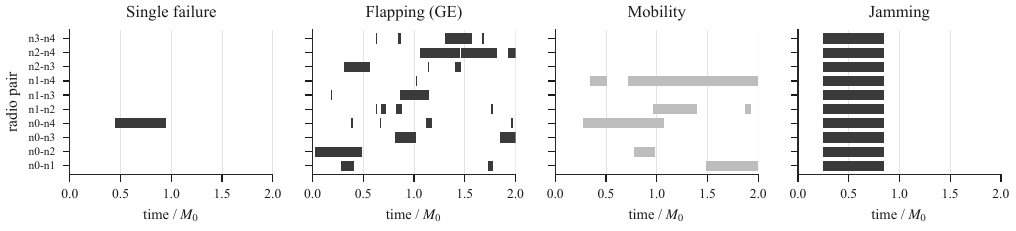}
    \caption{One instance's disruption traces, one panel per process: link state over time for every radio pair (dark: down; light: degraded; first $2\,M_0$ shown). The correlation structures differ on purpose: an isolated outage; memoryless flapping, independent across links; mobility's slow degradations that heal as nodes return; and jamming's blackout, severing every link at one instant.}
    \label{fig:traces}
\end{figure*}

All disruption processes compile to the same artifact, a trace of timed link events, so processes with very different structure run under identical workloads and policies; what distinguishes them is their correlation structure, precisely what we hypothesize the best repair scope depends on.
Fig.~\ref{fig:traces} shows one instance's trace under each process.
The four generators are a set of disruption processes rather than a unified adversary model: single failures, flapping, and mobility represent non-adversarial platform dynamics, while jamming is adversarial and attacks communication only.
In every case, the scheduler observes link-state events causally as they occur, with no knowledge of durations or recovery times; only the oracle sees the complete trace.
Disruptions affect connectivity and bandwidth but do not compromise nodes, alter task execution, falsify scheduler state, or attack the control plane, an assumption we return to in the conclusion.

\begin{figure*}[t]
    \centering
    \includegraphics[width=0.77\textwidth]{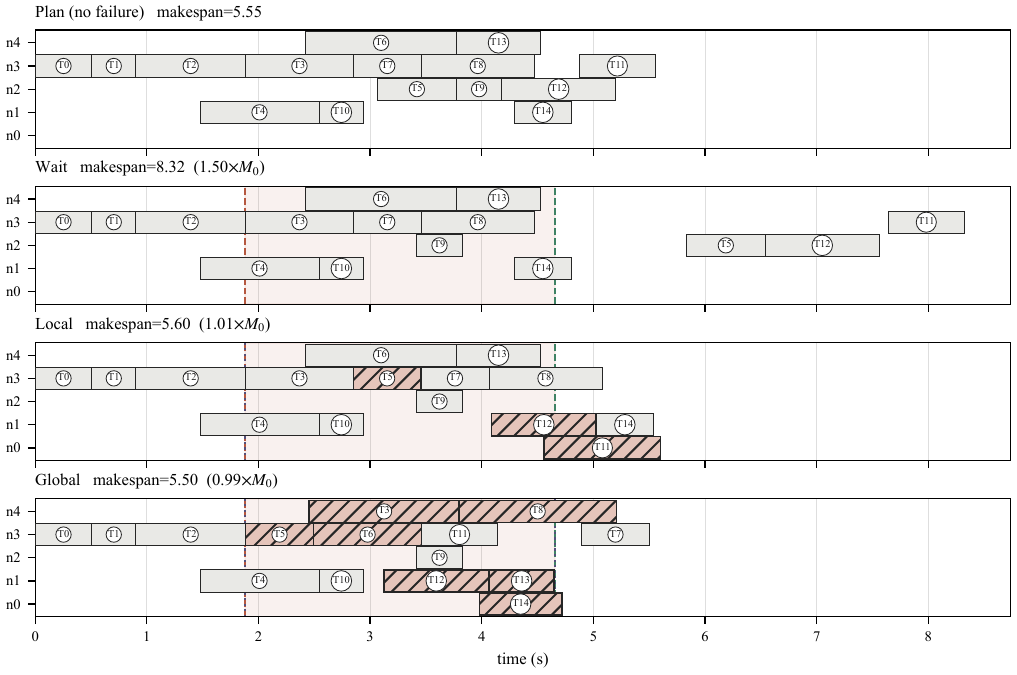}
    \caption{One episode, observed. The radio pair n2--n3 fails at $t{=}1.88$ (dashed; shaded until recovery at $t{=}4.65$) just as the plan's largest transfer would cross it. \emph{Wait} strands the dependent chain T5$\to$T12$\to$T11 until recovery ($1.50\times M_0$). \emph{Local} repair (applied at the dotted line) moves only the reached tasks and their descendants (hatched) and finishes at $1.01\times M_0$; \emph{global} replans seven tasks and beats the original heuristic plan ($0.99\times M_0$). Rerouting (not shown) detours at equal bandwidth and matches the plan exactly; the oracle here coincides with global repair.}
    \label{fig:episode}
\end{figure*}

\textbf{Single failure} is the controlled-analysis case: one radio pair fails at a chosen time for a chosen duration.

\textbf{Flapping} is the memoryless baseline. Each radio pair independently alternates exponential up and down times, with a mean time between failures (MTBF) and a mean time to recovery (MTTR), the classic two-state Markov link model~\cite{gilbert1960, elliott1963}; failures arrive without warning, independently across links, and outages are short.

\textbf{Mobility} produces trending, temporally correlated outages: nodes follow random-waypoint trajectories~\cite{nayebi2012linklifetime}, and each pair's bandwidth follows inter-node distance through a stepped ramp: full within $r_\mathrm{full}$, dead beyond $r_\mathrm{max}$, degrading between.
A failure announces itself: degradation strictly precedes disconnection, and links recover as nodes re-approach.

\textbf{Jamming} is the adversarial case: an area jammer activates without warning, severs every radio pair in range at one instant, and recovers just as abruptly when it stops.
At this deployment's scale, the jam blankets the node cluster, so jamming is a total blackout, the fully correlated extreme the memoryless model cannot produce; targeting subsets of a larger network, in the adversarial-instance spirit of~\cite{coleman2024saga, khodabandehlou2025preemption}, is future work.

\section{Evaluation}
\label{sec:evaluation}

\subsection{Setup}
\label{sec:setup}

We implement the disruption semantics, the repair ladder, and the four disruption generators in ncsim~\cite{krishnamachari2026ncsim}, a discrete-event simulator for DAG scheduling on networked compute that delegates placement to SAGA's schedulers~\cite{coleman2024saga}.

Workloads span three families: synthetic structures (out-trees, in-trees, parallel chains) with seeded weight streams, the four RIoTBench IoT stream-processing pipelines (ETL, STATS, TRAIN, PRED)~\cite{shukla2017riotbench}, and three WfCommons scientific workflows (Epigenomics, Montage, Seismology) drawn from distributions fitted to real execution traces~\cite{coleman2023wfchef}, with 45--102 tasks.
Ten instances of each of the ten structures run on five-node networks with heterogeneous compute speeds and full-mesh radio links.
Link bandwidths are set by the communication-to-computation ratio (CCR), the mean data transfer time divided by the mean task execution time: low-CCR instances are compute-bound, high-CCR ones communication-bound.
Each instance runs at five CCR values (0.2, 0.5, 1, 2, 5), yielding 500 instance variants; the aggregates pool across the grid unless a figure explicitly resolves CCR.

Each instance is normalized by its undisrupted makespan $M_0$ and by its oracle makespan.
Disruption processes are parameterized relative to $M_0$ so every instance is disrupted comparably: the single failure lasts $0.5 M_0$; flapping uses MTBF $0.75 M_0$ and MTTR $0.08 M_0$; mobility moves half the nodes at speeds that cross the deployment area in about $M_0$; the jammer is active for $0.6 M_0$.
The single failure is exhaustive rather than sampled: every radio pair fails at each of nine start times ($0.1$--$0.9\,M_0$), 90 cells per variant.
Failures on idle links cost nothing under any policy, so single-failure aggregates condition on the \emph{biting} cells, where waiting's makespan exceeds $M_0$ (19\% of cells).
The three stochastic processes remain pooled, a quiet draw being part of the threat model.
They bite in 71\% (jamming), 42\% (flapping), and 16\% (mobility) of variants.
Repair cost $\delta$ in Eq.~\eqref{eq:repair-cost} is likewise a fraction of $M_0$ per moved task, swept from 0 to 0.5.
We report the geometric mean of makespan over the oracle (ratios) across all completed instances.

\subsection{Results}
\label{sec:results}

\begin{figure*}[t]
    \centering
    \includegraphics[width=\textwidth]{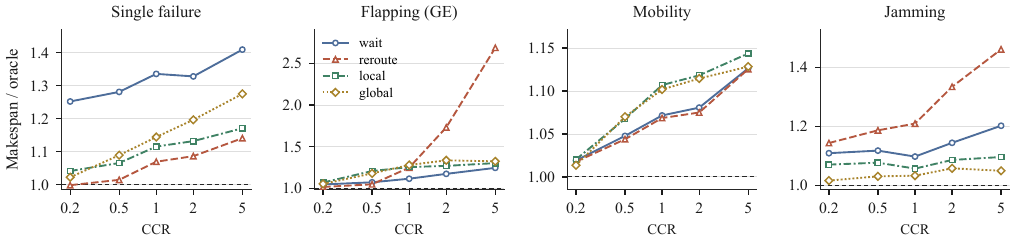}
    \caption{Makespan relative to the per-variant oracle for each repair policy under each disruption process, at zero repair cost, resolved by CCR (geometric means; per-panel vertical scales; the single failure uses the biting cells of the exhaustive pair--time sweep). Communication-bound variants amplify every verdict, except under flapping, where the winner flips: rerouting is best at CCR $\leq$ 0.5 and collapses to $2.7\times$ the oracle at CCR 5, where waiting wins.}
    \label{fig:ladder}
\end{figure*}

\begin{figure*}[t]
    \centering
    \includegraphics[width=0.86\textwidth]{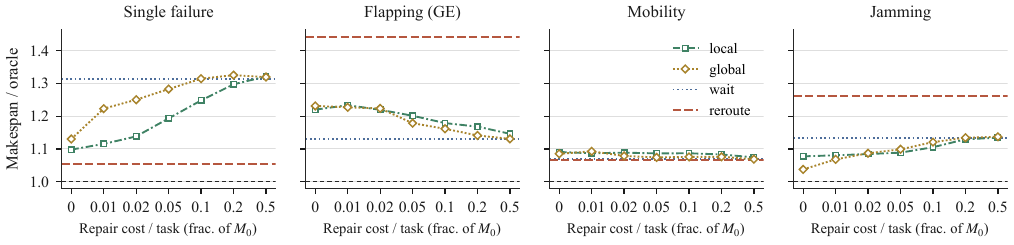}
    \caption{Makespan relative to the oracle as the per-moved-task repair cost $\delta$ grows (fractions of $M_0$; non-uniform grid). Wait and reroute never move tasks and appear as cost-independent references. Free repair favors the largest scope; charging for it erodes the advantage, and by $\delta = 0.2$--$0.5$ both scopes have converged onto the waiting line under every process.}
    \label{fig:cost}
\end{figure*}

\begin{table}[t]
    \caption{Geometric-mean makespan over the oracle by workload family (zero repair cost; best policy bold).}
    \label{tab:family}
    \centering
    \footnotesize
    \setlength{\tabcolsep}{4pt}
    \begin{tabular}{llcccc}
\toprule
Family & Disruption & Wait & Reroute & Local & Global \\
\midrule
RIoTBench & Single failure & 1.35 & \textbf{1.03} & 1.08 & 1.09 \\
 & Flapping (GE) & 1.06 & 1.18 & 1.05 & \textbf{1.04} \\
 & Mobility & 1.03 & 1.03 & 1.03 & \textbf{1.01} \\
 & Jamming & 1.11 & 1.14 & 1.05 & \textbf{1.04} \\
\midrule
Synthetic & Single failure & 1.36 & \textbf{1.04} & 1.09 & 1.07 \\
 & Flapping (GE) & 1.14 & 1.16 & 1.16 & \textbf{1.13} \\
 & Mobility & 1.08 & \textbf{1.07} & 1.08 & \textbf{1.07} \\
 & Jamming & 1.18 & 1.23 & 1.08 & \textbf{1.03} \\
\midrule
WfCommons & Single failure & 1.29 & \textbf{1.06} & 1.10 & 1.17 \\
 & Flapping (GE) & \textbf{1.23} & 2.33 & 1.58 & 1.68 \\
 & Mobility & 1.11 & \textbf{1.10} & 1.18 & 1.21 \\
 & Jamming & 1.13 & 1.48 & 1.12 & \textbf{1.04} \\
\bottomrule
\end{tabular}

\end{table}

Fig.~\ref{fig:episode} shows a single observable episode before any aggregation: a targeted failure on a loaded link and the four outcomes the ladder produces. Waiting strands a dependent chain for the entire outage, local repair moves three tasks, and global reshuffles seven, landing below the original plan.
Fig.~\ref{fig:ladder} then shows the ladder resolved by CCR, Table~\ref{tab:family} the per-family breakdown, and Fig.~\ref{fig:cost} the effect of charging for repair.
Every run in every cell completed its workload, as these disruption processes are transient.

\subsubsection{Where and when a failure bites}
For one representative instance at CCR 1, the exhaustive sweep bites in 22 of 90 (pair, time) cells, exactly where planned traffic meets the outage window, and costs nothing elsewhere.
This is the basis for the biting-cell conditioning above.
The cells also cleanly separate the mechanisms.
Rerouting erases nearly all the damage a detour can, and local repair caps the worst cells that waiting pays dearly for.
Part of global repair's advantage is re-optimization. Scheduling against any event also repairs the heuristic's original placement, occasionally finishing below $M_0$.

\subsubsection{The best scope tracks the disruption structure}
Under \emph{jamming} (abrupt, fully correlated, every link at once), repair pays.
Global repair comes within 4\% of the oracle and local within 8\%, while waiting gives up 13\% and rerouting 26\%: a blackout leaves nothing to detour onto during the outage, and at its edges rerouting's retries latch onto long multi-hop paths that occupy several radio links at once.
Under \emph{flapping}, the ordering inverts in the pooled aggregate, and waiting (1.13) beats both rescheduling policies and rerouting.
Memoryless short outages are gone before a repair's consequences are, so every reaction chases a link state that no longer holds, the scheduling analog of why BGP dampens route flapping~\cite{rfc2439}.
The \emph{single failure} is rerouting's case, exactly what a detour handles outright: on the cells where the failure bites, a detour nearly erases it (1.05) and local repair stays close, while global pays for its ambition on the largest workflows and waiting pays 1.31.
\emph{Mobility} refutes our own hypothesis: we expected its trending, warned failures to favor local repair, but waiting and rerouting beat both repair scopes.
Mobility outages self-heal as departed nodes return, so a placement that waits is a placement already repaired.
Reacting merely moves tasks toward nodes that are themselves about to leave.

\subsubsection{The communication regime amplifies, and can flip, the verdict}
Fig.~\ref{fig:ladder} resolves the ladder by CCR.
Under the three stochastic processes, CCR 0.2 barely separates the policies: when transfers are cheap, a link outage has little makespan to destroy, and any response is roughly free.
The biting single failures separate at every CCR, with waiting paying 25--41\% throughout while a detour stays near the oracle.
As CCR grows, every ordering above amplifies rather than reverses, with one exception: under flapping, the winner \emph{flips}.
Rerouting is the best policy up to CCR 0.5 yet collapses to $2.7\times$ the oracle at CCR 5, where waiting wins.
The mechanism is the abort-and-retransmit semantics: a compute-bound transfer completes its detour before the next flap. At the same time, a communication-bound one lives on the air long enough to be aborted repeatedly, each abort discarding the whole transfer.
Jamming's verdict, by contrast, is regime-independent, with global repair leading at every CCR.

\subsubsection{Workload scale shifts the optimum down the ladder}
Table~\ref{tab:family} shows the aggregate hides a strong workload effect.
On the small RIoTBench pipelines ($\sim$10 tasks), global repair leads under every stochastic process, even flapping: replanning ten tasks is too small an action to backfire.
On the larger WfCommons workflows (45--102 tasks), the picture reverses: under flapping, waiting decisively beats every reaction.
With many pending tasks, every reaction reshuffles a large schedule and pays for it on links that already recovered.
The synthetic structures sit between, with global repair best or tied under the stochastic processes, jamming most starkly.

\subsubsection{Repair latency erodes large scopes first}
Fig.~\ref{fig:cost} shows repair cost as an axis rather than an assumption.
At $\delta = 0$, global repair leads wherever repair helps.
By $\delta = 0.02 M_0$ per moved task, local repair matches or beats it under both the single failure and jamming, since global's larger plans pay sharply at the first price step.
By $\delta = 0.1 M_0$ the jamming advantage is a sliver, and by $\delta = 0.2$--$0.5$ both scopes converge onto the waiting line under every process.
A plan that arrives after the disruption has passed never takes effect, so expensive repair degenerates into waiting.
Under flapping and mobility, no price makes repair worthwhile.
Interestingly, under flapping, \emph{expensive} repair is milder than cheap repair, because long decision latencies allow newer plans to supersede pending ones, inadvertently damping churn.

\subsubsection{The design ladder}
Together, the findings distill into a single design ladder.
Do nothing when disruptions are short or self-healing.
Reroute around isolated persistent failures.
Repair locally when the disruption is localized, repair is priced, or the workload is large.
Repair globally when outages are broad, correlated, and persistent.
Rising repair costs or communication intensity push every case toward a smaller scope.

\section{Conclusion}
\label{sec:conclusion}

When the network moves under a task graph schedule, how much of the schedule should move with it?
We made repair scope an explicit, separately priced decision, comparing four policies (wait, reroute, local repair, global repair) under disruption processes whose correlation structures were deliberately different.
The data supports a design ladder rather than a winner: match the repair scope to the disruption's correlation structure, persistence, and price, and when in doubt react less, since every reaction chases a network state that may no longer hold.
Larger workflows and communication-bound regimes both amplify the case for restraint.
Two limitations remain and point at future work.
First, the control plane: our scheduler observes link state. It disseminates plans out of band, an assumption strongest exactly where global repair wins, since a blackout severs the data plane in both directions.
The cost sweep bounds how much control latency the advantage can withstand, but feasibility during a blackout requires a separate control channel or decentralized repair in the style of Jupiter~\cite{ghosh2021jupiter}.
Second, warning: our mobility results show warning alone is not enough, but proactive repair \emph{before} a forecast disconnection remains open.

\section*{Acknowledgment}
This work was supported in part by ARL under Cooperative Agreement W911NF-17-2-0196.

\bibliographystyle{IEEEtran}
\bibliography{references}

\begin{thebibliography}{10}
\providecommand{\url}[1]{#1}
\csname url@samestyle\endcsname
\providecommand{\newblock}{\relax}
\providecommand{\bibinfo}[2]{#2}
\providecommand{\BIBentrySTDinterwordspacing}{\spaceskip=0pt\relax}
\providecommand{\BIBentryALTinterwordstretchfactor}{4}
\providecommand{\BIBentryALTinterwordspacing}{\spaceskip=\fontdimen2\font plus
\BIBentryALTinterwordstretchfactor\fontdimen3\font minus
  \fontdimen4\font\relax}
\providecommand{\BIBforeignlanguage}[2]{{%
\expandafter\ifx\csname l@#1\endcsname\relax
\typeout{** WARNING: IEEEtran.bst: No hyphenation pattern has been}%
\typeout{** loaded for the language `#1'. Using the pattern for}%
\typeout{** the default language instead.}%
\else
\language=\csname l@#1\endcsname
\fi
#2}}
\providecommand{\BIBdecl}{\relax}
\BIBdecl

\bibitem{shukla2017riotbench}
A.~Shukla, S.~Chaturvedi, and Y.~Simmhan, ``{RIoTBench}: An {IoT} benchmark for
  distributed stream processing systems,'' \emph{Concurrency and Computation:
  Practice and Experience}, vol.~29, no.~21, p. e4257, 2017.

\bibitem{ghosh2021jupiter}
P.~Ghosh, Q.~Nguyen, P.~K. Sakulkar, J.~A. Tran, A.~Knezevic, J.~Wang, Z.~Lin,
  B.~Krishnamachari, M.~Annavaram, and S.~Avestimehr, ``Jupiter: A networked
  computing architecture,'' in \emph{Proc. 14th IEEE/ACM International
  Conference on Utility and Cloud Computing Companion}, 2021, pp. 1--8.

\bibitem{topcuoglu2002heft}
H.~Topcuoglu, S.~Hariri, and M.-Y. Wu, ``Performance-effective and
  low-complexity task scheduling for heterogeneous computing,'' \emph{IEEE
  Transactions on Parallel and Distributed Systems}, vol.~13, no.~3, pp.
  260--274, 2002.

\bibitem{khodabandehlou2025preemption}
M.~Khodabandehlou, J.~Coleman, N.~Suri, and B.~Krishnamachari, ``Studying the
  effect of schedule preemption on dynamic task graph scheduling,'' in
  \emph{Proc. IEEE Military Communications Conference (MILCOM)}, 2025.

\bibitem{krishnamachari2026ncsim}
B.~Krishnamachari, M.~Gutierrez, and J.~Coleman, ``ncsim: A lightweight
  simulator for networked edge computing with wireless interference modeling,''
  \emph{arXiv preprint arXiv:2605.01094}, 2026.

\bibitem{coleman2023wfchef}
T.~Coleman, H.~Casanova, and R.~Ferreira~da Silva, ``Automated generation of
  scientific workflow generators with {WfChef},'' \emph{Future Generation
  Computer Systems}, vol. 147, pp. 16--29, 2023.

\bibitem{setlur2020faulttolerant}
A.~R. Setlur, S.~J. Nirmala, H.~Singh, and S.~Khoriya, ``An efficient fault
  tolerant workflow scheduling approach using replication heuristics and
  checkpointing in the cloud,'' \emph{Journal of Parallel and Distributed
  Computing}, vol. 136, pp. 14--28, 2020.

\bibitem{mei2015ftdr}
J.~Mei, K.~Li, X.~Zhou, and K.~Li, ``Fault-tolerant dynamic rescheduling for
  heterogeneous computing systems,'' \emph{Journal of Grid Computing}, vol.~13,
  pp. 507--525, 2015.

\bibitem{sakellariou2004rescheduling}
R.~Sakellariou and H.~Zhao, ``A low-cost rescheduling policy for efficient
  mapping of workflows on grid systems,'' \emph{Scientific Programming},
  vol.~12, no.~4, pp. 253--262, 2004.

\bibitem{chamorro2025online}
J.~Chamorro, G.~Twigg-Ho, J.~Coleman, T.~Coleman, B.~Krishnamachari, and
  M.~Khodabandehlou, ``Adapting classic scheduling heuristics for online
  execution under uncertainty,'' in \emph{Proc. SC Workshops of the
  International Conference for High Performance Computing, Networking, Storage
  and Analysis (WORKS)}, 2025.

\bibitem{cai2021datr}
L.~Cai, X.~Wei, C.~Xing, X.~Zou, G.~Zhang, and X.~Wang, ``Failure-resilient
  {DAG} task scheduling in edge computing,'' \emph{Computer Networks}, vol.
  198, p. 108361, 2021.

\bibitem{khodabandehlou2025sensys}
M.~Khodabandehlou, J.~Coleman, and B.~Krishnamachari, ``Scheduling dynamic
  {IoT} task graphs,'' in \emph{Proc. 23rd ACM Conference on Embedded Networked
  Sensor Systems (SenSys)}, 2025, pp. 624--625.

\bibitem{perkins1999aodv}
C.~E. Perkins and E.~M. Royer, ``Ad-hoc on-demand distance vector routing,'' in
  \emph{Proc. 2nd IEEE Workshop on Mobile Computing Systems and Applications
  (WMCSA)}, 1999, pp. 90--100.

\bibitem{rfc4090}
P.~Pan, G.~Swallow, and A.~Atlas, ``Fast reroute extensions to {RSVP-TE} for
  {LSP} tunnels,'' IETF, Tech. Rep. RFC 4090, 2005.

\bibitem{rfc2439}
C.~Villamizar, R.~Chandra, and R.~Govindan, ``{BGP} route flap damping,'' IETF,
  Tech. Rep. RFC 2439, 1998.

\bibitem{pozo2021shp}
F.~Pozo, G.~Rodr{\'i}guez-Navas, and H.~Hansson, ``Self-healing protocol:
  Repairing schedules online after link failures in time-triggered networks,''
  in \emph{Proc. 51st IEEE/IFIP International Conference on Dependable Systems
  and Networks (DSN)}, 2021.

\bibitem{asemian2025jamming}
G.~Asemian, M.~Amini, and B.~Kantarci, ``Reliable task offloading in {MEC}
  through transmission diversity and jamming-aware scheduling,'' in \emph{Proc.
  International Conference on Network of the Future (NoF)}, 2025.

\bibitem{gilbert1960}
E.~N. Gilbert, ``Capacity of a burst-noise channel,'' \emph{Bell System
  Technical Journal}, vol.~39, no.~5, pp. 1253--1265, 1960.

\bibitem{elliott1963}
E.~O. Elliott, ``Estimates of error rates for codes on burst-noise channels,''
  \emph{Bell System Technical Journal}, vol.~42, no.~5, pp. 1977--1997, 1963.

\bibitem{nayebi2012linklifetime}
A.~Nayebi and H.~Sarbazi-Azad, ``Analysis of link lifetime in wireless mobile
  networks,'' \emph{Ad Hoc Networks}, vol.~10, no.~7, pp. 1221--1237, 2012.

\bibitem{coleman2024saga}
J.~Coleman and B.~Krishnamachari, ``Comparing task graph scheduling algorithms:
  An adversarial approach,'' \emph{arXiv preprint arXiv:2403.07120}, 2024.

\end{thebibliography}

\end{document}